\documentclass[letterpaper]{article}

\usepackage[T1]{fontenc}

\usepackage{geometry}
\usepackage{setspace}
\usepackage{amsmath}

\usepackage[style = chem-acs]{biblatex}
\usepackage{graphicx}
\usepackage{float}
\usepackage{url}
\newfloat{scheme}{htbp}{los}
\floatname{scheme}{Scheme}
\floatname{chart}{Chart}
\newfloat{graph}{htbp}{loh}

\usepackage{chemformula} 
\usepackage[version = 4]{mhchem} 

\usepackage{authblk}
\author[1,2,$\dagger$]{Chenghan Li}
\author[1,2]{Garnet Kin-Lic Chan*}
\affil[1]{Division of Chemistry and Chemical Engineering, California Institute of Technology, Pasadena, CA 91125, USA}
\affil[2]{Marcus Center for Theoretical Chemistry, California Institute of Technology, Pasadena, CA 91125, USA}

\newcommand{\pd}{\partial}
\newcommand{\pdpd}[2]{{\frac{\pd #1}{\pd #2}}}
\newcommand{\ncon}{{m}}

\newcommand{\mEh}{{\text{m}E_\text{h}}}

\title{A Constrained Coupled Cluster Approach for Strongly Correlated Electrons}
\date{*Email: gkc1000@gmail.com}

\begin{document}

\maketitle
\begingroup\renewcommand\thefootnote{$\dagger$}\footnotetext{Current address: Department of Molecular Biology and Biochemistry, University of California, Irvine, California 92697, United States}\endgroup

\begin{abstract}
    We introduce constrained coupled cluster (cCC), a general framework for imposing arbitrary physical constraints on conventional single-reference coupled cluster amplitudes without rederiving or replacing the underlying coupled cluster residual equations.
    As a first application, we derive a constraint from the spin-symmetry requirement for an exact wavefunction and demonstrate the resulting approach on strongly correlated problems. 
    Across benchmark Hubbard models, nitrogen and chromium bond breaking, and iron--sulfur cluster spin gaps, the constraint strongly suppresses spin contamination, repairs qualitative failures of conventional coupled cluster, and often brings CCSDT, and in several cases CCSD, into quantitative agreement with exact diagonalization or high-accuracy density matrix renormalization group calculations. The derived constraint adds only a mean-field cost contraction to the coupled cluster calculation and is compatible with any cluster truncation. These results identify numerical constraints as a promising route for extending single-reference coupled cluster machinery into strongly correlated regimes conventionally associated with more complicated electronic structure methods.
\end{abstract}

\section*{Keywords}

Coupled cluster theory, strong correlation, spin symmetry, transition-metal chemistry

%

\section{Introduction}

Coupled-cluster (CC) theory is among the most successful electronic-structure methods for weakly to moderately correlated systems.\cite{Bartlett2007} However, there are many problems that are not well-captured by a single-determinant starting point and where several determinants acquire comparable weights, as found in bond dissociation, polyradicals, transition-metal clusters, and strongly correlated lattice models.\cite{chan2024spiers} Developing a CC method that preserves the machinery and favorable structure of conventional CC while describing multi-determinant chemical regimes has therefore remained a long-standing goal.



In this work, we introduce constrained coupled cluster (cCC) theory, in which user-defined constraints are imposed directly on conventional single-reference CC amplitudes. The exponential ansatz, energy expression, and residual contractions are retained, and only the convergence condition is changed to numerically satisfy the imposed constraints. A new physical constraint can therefore be added as a modular layer around existing CC machinery, without a constraint-specific rederivation of its diagrammatic algebra.

As an application of this framework within a strong correlation setting, we consider the problem of spin symmetry within open-shell coupled cluster theory. There is a long history of attempts to exactly, or approximately, ensure the spin symmetry of coupled cluster solutions in strong correlation settings.
Existing approaches broadly either encode symmetry into the reference, excitation operators, or ansatz,\cite{Janssen1991,Li1993,nooijen_general_1996,Herrmann2020} or restore it by projecting a broken-symmetry wavefunction onto its symmetric component.\cite{Qiu2017,Qiu2018,song2024combining,song2026multisymmetryprojectedcoupledcluster}
For example, methods that encode spin symmetry in the reference require a
multiconfigurational ansatz, and include various flavours of multireference CC\cite{Rittby1991,Lyakh2012,Balkova1992,Datta2009} and spin-flip equation-of-motion CC.\cite{Krylov2001,Casanova2020} In the first case, this gives up the convenience of the single-reference CC methodology, while the second is limited (in practice) to a small number of coupled spins. Methods that use symmetry-adapted excitation operators work with the unitary-group algebra and spin-complete manifolds of spatial substitution operators.\cite{Li1993,li1994automation,Herrmann2020,Herrmann2022} However, spin completeness can require spectator substitutions beyond ordinary particle--hole excitations, and these generalized substitutions contract with one another, preventing the usual low-order termination of the Baker--Campbell--Hausdorff expansion.\cite{nooijen_general_1996} Finally, spin projection techniques allow for the use of broken-symmetry mean-field determinants while (approximately) restoring spin symmetry, but with an increase in cost, which in principle increases with the number of spin components, and which in practice can be an order of magnitude or more (e.g., due to the number of quadrature points).\cite{Qiu2017,Qiu2018,song2024combining,song2026multisymmetryprojectedcoupledcluster} 

The current situation leaves an important gap for a method that works directly with a symmetry-broken determinant, extends to many open shells, and still retains the conventional cost and infrastructure of single-reference CC. 
We fill this gap by introducing spin constraints on the coupled cluster equations as a way to restore spin symmetry. 
We derive inexpensive spin constraints from the exact highest-weight relation $S_+|\Psi\rangle=0$, projected into $S_+$-closed configuration-interaction (CI) subspaces. Because the constraints act on CI coefficients generated by $e^T|0\rangle$, they can be combined with unrestricted orbitals and symmetry-broken references and applied without change to many-open-shell low-spin states. The subspace constraints are automatically satisfied by an exact wavefunction, so cCC recovers the exact solution in the untruncated cluster limit, unlike the linearized amplitude relations used in partially spin-adapted CC.\cite{Janssen1991,knowles_coupled_1993,Szalay1997}


Our numerical results show that the spin-constrained CC equations significantly improve on standard broken-symmetry unrestricted CC calculations with little increase in cost. Across a series of test problems that include Hubbard models, molecular bond breaking, and iron--sulfur spin gaps, a simple spin constraint in the CI singles and doubles (CISD) space strongly reduces spin contamination and reduces the energy errors of conventional broken-symmetry CC, in some cases by an order of magnitude, while the additional $\mathcal O(N^4)$ contractions introduced by the spin constraint only marginally increase the cost. In particular, our finding is that the truncated CISD spin constraint should always be applied when carrying out an unrestricted CC calculation, as it always improves the result. More generally, our work positions spin-constrained CC as a first product of a broader framework: using numerical physical constraints to carry the established strengths of single-reference CC into more complex regimes of electronic structure.

\section{Theory}

\subsection{Coupled Cluster with Constraints}



The amplitudes in standard CC are obtained by solving the residual equations for a reference determinant $|0\rangle$:
\begin{align}
    R_\mu(t) &= \langle\mu|\bar{H}(t)|0\rangle = 0,
    \label{eq:cc_residual}
\end{align}
Here $\bar{H}=e^{-T}H e^T$ is the similarity-transformed Hamiltonian. The cluster operator $T=\sum_\mu t_\mu\tau_\mu$ consists of particle--hole excitation operators $\tau_\mu$ with amplitudes $t_\mu$.

We now introduce $\ncon$ differentiable constraints on the amplitudes,
\begin{align}
    C_A(t)=0.
    \label{eq:ccc_constr}
\end{align}
Here and below, $A=1,\ldots,\ncon$. We denote their Jacobian by
\begin{align}
    G_{A\mu}(t)=\pdpd{C_A(t)}{t_\mu}.
    \label{eq:constraint_jacobian}
\end{align}
The constrained CC (cCC) equations are defined as
\begin{align}
    R_\mu(t) &= \sum_A \eta_A G_{A\mu}(t),
    \label{eq:ccc_residual}\\
    C_A(t) &=0,
    \nonumber
\end{align}
where the multipliers $\eta_A$ are solved for together with the cluster amplitudes. The CC energy is still evaluated in the regular way, i.e., $E^\text{CC}=\langle0|\bar{H}|0\rangle$. Because of the constraints, the residuals $R_\mu(t)$ no longer vanish, rather 
Eq.~\ref{eq:ccc_residual} requires only that the projection of the residual vector onto the tangent space of the constraint manifold vanishes. For locally independent constraints, Eqs.~\ref{eq:ccc_constr} and \ref{eq:ccc_residual} provide $|T|+\ncon$ equations for the same number of unknowns, $t$ and $\eta$.

Clearly, ordinary CC is recovered when no constraints are present. As a simple example of a constrained theory, we can consider tailored CC\cite{Kinoshita2005TailoredCC}. The tailored-CC amplitude equations\cite{Kinoshita2005TailoredCC} are recovered by fixing the internal active-space amplitudes with constraints of the form $C_p(t)=t_p-t_p^0$, where $p$ labels an active-space amplitude and $t_p^0$ is usually obtained by cluster decomposition of a CASCI wavefunction. For these linear constraints, $G_{p\mu}=\delta_{p\mu}$, so Eq.~\ref{eq:ccc_residual} becomes $R_\mu(t) = 0$ for $\mu$ outside of the active space and $R_p = \eta_p$ within the active space; this defines the multipliers $\eta_p$ and places no additional conditions on the internal amplitudes (which are then entirely fixed by the constraint). Then only the external amplitudes are determined from the CC residual equations, as expected for tailored CC.

Computationally, Eqs.~\ref{eq:ccc_constr} and \ref{eq:ccc_residual} can be solved by correcting a conventional diagonally preconditioned CC iteration. Let $\widetilde t_\mu$ be the ordinary CC update and $D_\mu$ the diagonal orbital-energy approximation to the CC Jacobian. We use
\begin{align}
    t_\mu^\mathrm{new}
    =\widetilde t_\mu
    +\frac{1}{D_\mu}\sum_A\eta_A G_{A\mu}(t).
    \label{eq:ccc_update_t}
\end{align}
Requiring the updated amplitudes to satisfy the constraints to first order gives the constraint-space system
\begin{align}
    \sum_B M_{AB}\eta_B &= b_A,
    \label{eq:eta_linear_system}\\
    M_{AB} &= \sum_\mu G_{A\mu}D_\mu^{-1}G_{B\mu},
    \label{eq:constraint_metric}\\
    b_A &= -C_A(t)-\sum_\mu G_{A\mu}(\widetilde t_\mu-t_\mu).
    \label{eq:eta_rhs}
\end{align}
Recomputing the constraints and their derivatives after each amplitude update recovers the nonlinear cCC equations at convergence. More numerical details are given in the SI.

\subsection{Lagrangian and Response}

To obtain energy derivatives, we introduce the cCC Lagrangian
\begin{align}
    \mathcal{L}(t,\eta,\lambda,\kappa)
    ={}& E^\text{CC}(t)
    + \sum_\mu \lambda_\mu
    \left[R_\mu(t)-\sum_A\eta_A G_{A\mu}(t)\right]
    + \sum_A\kappa_A C_A(t).
    \label{eq:ccc_l}
\end{align}
Stationarity with respect to $\lambda$ and $\kappa$ recovers the cCC equations. Stationarity with respect to $\eta$ gives
\begin{align}
    \sum_\mu G_{A\mu}\lambda_\mu=0,
    \label{eq:ccc_leq1}
\end{align}
for $A=1,\ldots,\ncon$.
For the remaining equations, define
\begin{align}
    e_\nu &= \pdpd{E^\text{CC}}{t_\nu}
    =\langle0|[\bar H,\tau_\nu]|0\rangle,
    \label{eq:energy_amplitude_derivative}\\
    J_{\mu\nu} &= \pdpd{R_\mu}{t_\nu}
    =\langle\mu|[\bar H,\tau_\nu]|0\rangle,
    \label{eq:cc_jacobian}\\
    \mathcal{C}^{A}_{\mu\nu}
    &=\frac{\pd^2 C_A}{\pd t_\mu\pd t_\nu}.
    \label{eq:constraint_hessian}
\end{align}
Stationarity with respect to $t_\nu$ then gives
\begin{align}
    e_\nu
    +\sum_\mu\lambda_\mu
    \left(J_{\mu\nu}-\sum_A\eta_A\mathcal{C}^{A}_{\mu\nu}\right)
    +\sum_A\kappa_A G_{A\nu}=0.
    \label{eq:ccc_leq2}
\end{align}
Equations~\ref{eq:ccc_leq1} and \ref{eq:ccc_leq2} form a linear system for the response multipliers $\lambda$ and $\kappa$. With
\begin{align}
    K_{\mu\nu}=J_{\mu\nu}-\sum_A\eta_A\mathcal{C}^{A}_{\mu\nu},
    \label{eq:effective_jacobian}
\end{align}
the system can be written compactly as
\begin{align}
    \begin{pmatrix}
        K^\mathrm{T} & G^\mathrm{T}\\
        G & 0
    \end{pmatrix}
    \begin{pmatrix}
        \lambda\\
        \kappa
    \end{pmatrix}
    =-
    \begin{pmatrix}
        e\\
        0
    \end{pmatrix}.
    \label{eq:ccc_lambda_block}
\end{align}
Thus, the usual CC $\Lambda$ equations are augmented by constraint derivatives and $\ncon$ additional multipliers. The resulting $t$ and $\lambda$ amplitudes from cCCSD can then be used in place of their ordinary CC counterparts. For example, below we use them to define a $\Lambda$-based perturbative triples correction\cite{lambdat}.

Let $\xi$ denote an external perturbation. Once the Lagrangian is stationary, the unrelaxed energy derivative (i.e., ignoring the mean-field orbital response) is its explicit derivative,
\begin{align}
    \frac{\mathrm{d}E}{\mathrm{d}\xi}
    ={}&\pdpd{\mathcal L}{\xi}
    = \langle0|\bar H^{(\xi)}|0\rangle
    +\sum_\mu\lambda_\mu\langle\mu|\bar H^{(\xi)}|0\rangle
    \nonumber\\
    &-\sum_{\mu A}\lambda_\mu\eta_A
      \pdpd{G_{A\mu}}{\xi}
    +\sum_A\kappa_A\pdpd{C_A}{\xi},
    \label{eq:ccc_response}
\end{align}
where $\bar H^{(\xi)}=e^{-T}(\pd H/\pd\xi)e^T$, with the amplitudes and orbitals held fixed in the partial derivative. When the constraints have no explicit dependence on $\xi$, the last two terms vanish, and Eq.~\ref{eq:ccc_response} has the same form as the ordinary CC response expression, evaluated with the constrained $t$ and $\lambda$ amplitudes.

In particular, the unrelaxed density matrices are obtained as the response with respect to the Hamiltonian matrix elements. The above shows that the one- and two-particle unrelaxed density matrices are given by the standard formulae
\begin{align}
    \gamma_{pq}
    ={}&\langle0|(1+\Lambda)e^{-T}a_p^\dagger a_qe^T|0\rangle,
    \label{eq:ccc_1rdm}\\
    \Gamma_{pq,rs}
    ={}&\langle0|(1+\Lambda)e^{-T}a_p^\dagger a_q^\dagger a_s a_re^T|0\rangle,
    \label{eq:ccc_2rdm}
\end{align}
with $\Lambda = \sum_\mu \lambda_\mu \tau_\mu^\dagger$.
We use them below to evaluate $\langle S^2\rangle$ as one diagnostic of spin contamination.

\subsection{Spin Symmetry Constraints}

We now implement spin symmetry constraints within this general framework. 
We consider a wave function with fixed $M_S$ and target its highest-weight component, $M_S=S$. 
Such a spin eigenstate must satisfy
\begin{align}
    S_+|\Psi\rangle=0.
    \label{eq:sp_exact}
\end{align}
For a determinant $|\mu\rangle$ in the $M_S$ sector, denote
\begin{align}
    c_\mu(t)=\langle \mu|e^T|0\rangle,
    \label{eq:cc_ci_coefficient}
\end{align}
and then projection of Eq.~\ref{eq:sp_exact} onto determinants $|\nu^+\rangle$ in the $M_S+1$ sector gives
\begin{align}
    \sum_\mu S^+_{\nu\mu}c_\mu(t)&=0,\nonumber\\
    S^+_{\nu\mu}&=\langle \nu^+|S_+|\mu\rangle.
    \label{eq:full_spin_constraint}
\end{align}

Eq.~\ref{eq:full_spin_constraint} is written in the full CI space. We now consider restricting it to a CI subspace. Let $P$ be an $M_S$-projector and $P^+$ an $(M_S+1)$-projector. They are $S_+$-closed when
\begin{align}
    P^+S_+(1-P)=0.
    \label{eq:sp_closed}
\end{align}
Equivalently, every $S_+$ preimage of a determinant in $P^+$ belongs to $P$. For each independent determinant $|\nu^+\rangle$ in $P^+$, we impose
\begin{align}
    C_\nu(t)
    &=\langle \nu^+|P^+S_+Pe^T|0\rangle
    \nonumber\\
    &=\sum_{\mu\in P}S^+_{\nu\mu}c_\mu(t)=0.
    \label{eq:projected_spin_constraint}
\end{align}
Importantly, any spin-pure state in the $S=M_S$ sector satisfies Eq.~\ref{eq:projected_spin_constraint} for every $S_+$-closed pair. Consequently, the full-cluster representation of the exact wavefunction satisfies the constraints in any truncated CI space, and thus any spin-constrained CC remains exact in the untruncated $T$ limit. 

%
The formulation does not require restricted orbitals, since the raising operator can be written in non-orthogonal $\alpha$ and $\beta$ orbitals (bars denote $\beta$ orbitals)
\begin{align}
     S_+
     =\sum_{pq}X_{p\bar{q}}
       a_{p}^\dagger a_{\bar{q}},
     \label{eq:sp_uhf}
\end{align}
with
\begin{align}
     X_{p\bar{q}}
     =\langle\phi_{p}|\phi_{\bar{q}}\rangle.
\end{align}
This permits the same projected constraints to be applied to both restricted and unrestricted references.


Because Eq.~\ref{eq:projected_spin_constraint} is projected into a subspace, it is a necessary but not sufficient condition for spin purity. Choosing a subspace selects a set of partial spin constraints. Here, we consider two types of $S_+$-closed subspaces. The first is a configuration state function (CSF) space. It contains all the determinants with fixed $M_S$ and a fixed set of doubly, and singly occupied open-shell spatial orbitals (with varying spin occupancy).
Applying Eq.~\ref{eq:projected_spin_constraint} then restricts the component of the wavefunction in this space to lie in the span of the CSFs with $S=M_S$. 

The second construction uses a truncated CI space. This construction is convenient to apply with symmetry-broken references. Let $P_k$ project onto all $M_S$ determinants whose excitation rank relative to $|0\rangle$ is at most $k$. For $k=2$, i.e., the CISD space, the projector
can be shown to be
\begin{align}
    P^+_2 = \sum_{ia} |_{\bar{i}}^{a}\rangle \langle_{\bar{i}}^{a}|,
\end{align}
i.e., all the $\beta$ to $\alpha$ single spin flip excitations from the reference. In the above and from now on, we let $i,j,k,l$ and $a,b,c,d$ denote occupied and virtual $\alpha$ orbitals, respectively, with barred indices denoting the corresponding $\beta$ spaces. 

We write down the CISD constraint in tensor contraction form. Using the conventional excitation phases, the CISD coefficients entering a constraint associated with $|_{\bar{i}}^{a}\rangle=a_a^\dagger a_{\bar{i}}|0\rangle$ are
\begin{align}
    c_j^a &= t_j^a,\nonumber\\
    c_{\bar{i}}^{\bar{b}}&=t_{\bar{i}}^{\bar{b}},\nonumber\\
    c_{j\bar{i}}^{a\bar{b}}
    &=t_{j\bar{i}}^{a\bar{b}}
      +t_j^a t_{\bar{i}}^{\bar{b}}.
    \label{eq:cisd_ci_tensors}
\end{align}
Eq.~\ref{eq:projected_spin_constraint} then becomes
\begin{align}
 C_{a\bar{i}}={}&X_{a\bar{i}}
 -\sum_jX_{j\bar{i}}t_j^a
 +\sum_{\bar b}X_{a\bar b}t_{\bar{i}}^{\bar b}
 \nonumber\\
 &-\sum_{j\bar b}X_{j\bar b}
 \left(t_{j\bar{i}}^{a\bar b}
       +t_j^a t_{\bar{i}}^{\bar b}\right)=0.
 \label{eq:cisd_spin_tensor}
\end{align}
For completeness, define the intermediates
\begin{align}
 A_{j\bar{i}}&=X_{j\bar{i}}
   +\sum_{\bar b}X_{j\bar b}t_{\bar{i}}^{\bar b},\nonumber\\
 B_{a\bar b}&=X_{a\bar b}
   -\sum_jX_{j\bar b}t_j^a.
 \label{eq:cisd_spin_intermediates}
\end{align}
The nonzero Jacobian blocks are
\begin{align}
 \frac{\partial C_{a\bar{i}}}{\partial t_k^c}
   &=-\delta_{ac}A_{k\bar{i}},\nonumber\\
 \frac{\partial C_{a\bar{i}}}
      {\partial t_{\bar l}^{\bar d}}
   &=\delta_{\bar{i}\bar l}B_{a\bar d},\nonumber\\
 \frac{\partial C_{a\bar{i}}}
      {\partial t_{k\bar l}^{c\bar d}}
   &=-\delta_{ac}\delta_{\bar{i}\bar l}X_{k\bar d}.
 \label{eq:cisd_spin_jacobian}
\end{align}
All other amplitude blocks vanish. The leading contraction in Eq.~\ref{eq:cisd_spin_tensor} costs
$\mathcal O(n_{o_\alpha}n_{o_\beta}n_{v_\alpha}n_{v_\beta})
=\mathcal O(N^4)$. The Jacobian contains only $\mathcal O(N^4)$ nonzero elements, so its matrix-vector products can be evaluated with the same leading cost. Because Eq.~\ref{eq:cisd_spin_tensor} is quadratic in the amplitudes, the Hessian contractions required by the response equations are also no more expensive.

For the CI singles, doubles, and triples (CISDT) space ($k=3$),
\begin{align}
    P^+_3 ={}& P^+_2
    + \sum_{ijab} |_{i\bar{j}}^{ab}\rangle
      \langle_{i\bar{j}}^{ab}|\nonumber\\
    &+ \sum_{ijab} |_{\bar{i}\bar{j}}^{a\bar{b}}\rangle
      \langle_{\bar{i}\bar{j}}^{a\bar{b}}|.
\end{align}
This space includes bare $\beta\rightarrow\alpha$ single spin flips and spin flips dressed with particle--hole excitations. The CISDT target space contains $\mathcal O(N^4)$ constraints. Evaluating these constraints and the associated Jacobian and Hessian actions scales as $\mathcal O(N^6)$. Thus, a CISDT constraint has the formal scaling of a CCSD residual evaluation and remains lower order than the $\mathcal O(N^8)$ CCSDT residual evaluation.

\section{Results and discussion}


\subsection{Hubbard Models}

\begin{figure}[htbp]
\centering
\includegraphics[width=\linewidth]{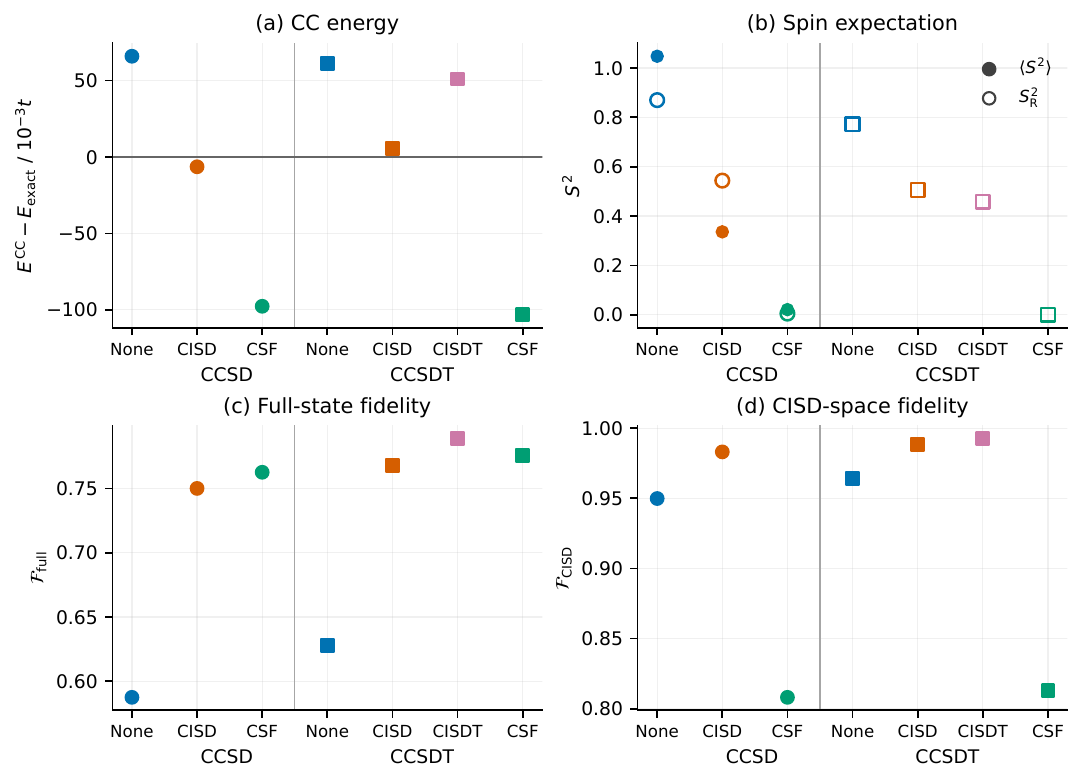}
\caption{Constraint-space comparison for the eight-site Hubbard chain at $U/t=16$ with a fixed N\'eel reference: (a) CC energy errors relative to exact diagonalization, (b) spin-squared expectation values, (c) full-space ket fidelities, and (d) normalized CISD-space fidelities. In (b), filled symbols denote the CCSD biorthogonal values $\langle S^2\rangle$ and open symbols denote the normalized ket values $S_R^2$; CCSDT biorthogonal values are unavailable. Circles and squares denote CCSD and CCSDT, respectively.}
\label{fig:hubbard_constraints}
\end{figure}

We first use an exactly solvable eight-site Hubbard model with open boundaries to analyze the behavior of cCC. We consider half filling at $U/t=16$. At this interaction strength, the site basis reasonably approximates the UHF orbitals. Starting from a fixed N\'eel determinant in the site basis therefore lets us compare the CSF (with fixed spatial orbitals) and truncated-CI constraints (for an `unrestricted' reference) on the same initial state.

We assess these spin constraints through constrained CC energies, spin diagnostics, and wavefunction fidelities, at the CCSD and CCSDT levels. Because we did not implement the $\Lambda$ equations at the CCSDT level, we report normalized spin-squared values from the CC ket, $|\Psi_R\rangle=e^T|0\rangle$,
\begin{align}
 S^2_{\mathrm R}&=\frac{\langle\Psi_R|S^2|\Psi_R\rangle}{\langle\Psi_R|\Psi_R\rangle}.
\end{align}
For CCSD, we additionally report the biorthogonal expectation $\langle S^2\rangle=\langle\Psi_L|S^2|\Psi_R\rangle$, with $\langle\Psi_L|=\langle0|(1+\Lambda)e^{-T}$, after solving the constrained response equations in Eq.~\ref{eq:ccc_lambda_block}. We also report the normalized full-space and CISD-space fidelities
\begin{align}
 \mathcal F_{\mathrm{full}}
 &=\frac{|\langle\Psi_\mathrm{exact}|\Psi_R\rangle|^2}
 {\langle\Psi_\mathrm{exact}|\Psi_\mathrm{exact}\rangle
  \langle\Psi_R|\Psi_R\rangle},\\
 \mathcal F_{\mathrm{CISD}}
 &=\frac{|\langle\Psi_\mathrm{exact}|P_2|\Psi_R\rangle|^2}
 {\langle\Psi_\mathrm{exact}|P_2|\Psi_\mathrm{exact}\rangle
  \langle\Psi_R|P_2|\Psi_R\rangle}.
\end{align}
The CISD fidelity is relevant to the energy since the CC energy expression for a two-body Hamiltonian depends only on the Hamiltonian matrix elements, reference, single-, and double-excitation CI coefficients.

Figure~\ref{fig:hubbard_constraints} compares these quantities for unconstrained, CSF-constrained, and CI-constrained CCSD and CCSDT. The CISD constraint reduces the conventional CC energy error by about a factor of ten for both CCSD and CCSDT. It also substantially reduces spin contamination according to both $S_R^2$ and $\langle S^2\rangle$, although enforcing spin symmetry in the CISD space is insufficient to make the full CC wavefunction spin pure. At half filling, the singlet ground state requires equal-magnitude N\'eel and anti-N\'eel coefficients. The anti-N\'eel determinant, obtained by flipping every spin in $|0\rangle$, lies beyond the CISD space for eight sites. It is included in the CSF space and is accessible through disconnected products of $T$. This allows the CSF constraint to remove the ket-state spin contamination almost completely at both CCSD and CCSDT levels (Figure~\ref{fig:hubbard_constraints}b).

However, although the CSF constraint is very effective at controlling the total spin contamination and improves the full-space fidelity, its CC energies are worse than the unconstrained values (Figure~\ref{fig:hubbard_constraints}a). 
The CSF constraint forces the anti-N\'eel CI coefficient to be of order one, which increases the cluster amplitudes substantially: $\max(|t_2|)$ rises from 0.53 in UCCSD to 0.98 in CSF-constrained CCSD. These large cluster amplitudes can be expected to create unphysical coefficients outside of the CSF space, and consequently, the greater spin purity does not ensure a more accurate CC energy.


A related behaviour is seen with CCSDT, when replacing the CISD constraint by CISDT.
This reduces the spin contamination and improves the fidelity.  
However, the CC energy error of CISDT-cCCSDT is larger than that of CISD-cCCSDT, although it is still better than that of the unconstrained CCSDT. It is clear that there must be a balance between restoring the spin purity and the symmetry breaking that improves the energy. Our results suggest
that, at the CCSD and CCSDT level, the CISD constraint provides the best balance between the two.

\begin{figure}[htbp]
\centering
\includegraphics[width=\linewidth]{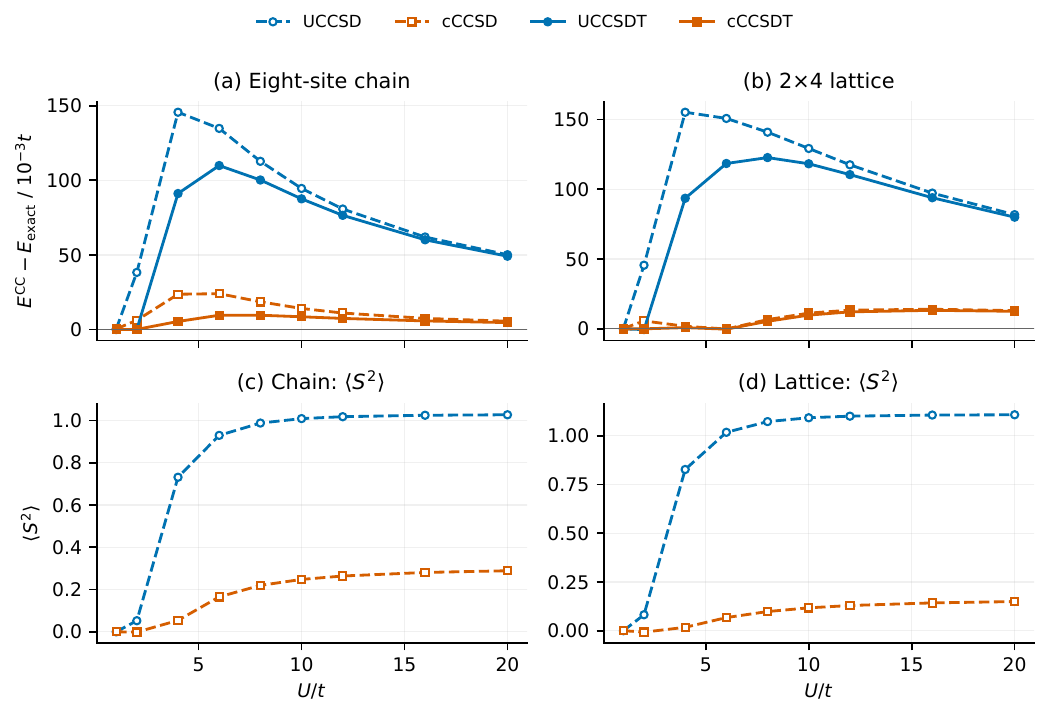}
\caption{Half-filled Hubbard models with UHF references and open boundaries. Panels (a,b) show CC energy errors relative to exact diagonalization for the chain and $2\times4$ lattice; panels (c,d) show $\langle S^2\rangle$ evaluated from the CCSD response density. The constrained methods use the CISD-space constraint. Lines connect the sampled interaction strengths. CCSDT values of $\langle S^2\rangle$ are unavailable and are omitted.}
\label{fig:hubbard_scan}
\end{figure}

To further examine the CISD-constrained cCC accuracy in a broader setting, we scan $U/t$ from 1 to 20 on an eight-site 1D Hubbard chain and a $2\times4$ 2D Hubbard lattice. The flexibility of the CISD constraint enables us to use a self-consistent UHF reference, which smoothly interpolates the band basis at the $U=0$ limit and the site basis at the $U=\infty$ limit. Figure~\ref{fig:hubbard_scan} shows the cCC energy error and the spin purity, compared to unconstrained CC. In the intermediate-correlation region around $U/t\approx6$, where the regular CC errors are near their broad maximum, CISD-cCC reduces the CCSD and CCSDT errors by about one order of magnitude on the chain, and by several hundredfold on the $2\times4$ lattice. Again, spin contamination is significantly suppressed, as shown by a $\sim5-10$ fold reduction in $\langle S^2\rangle$.

\subsection{Bond Breaking}

\begin{figure}[htbp]
\centering
\includegraphics[width=\linewidth]{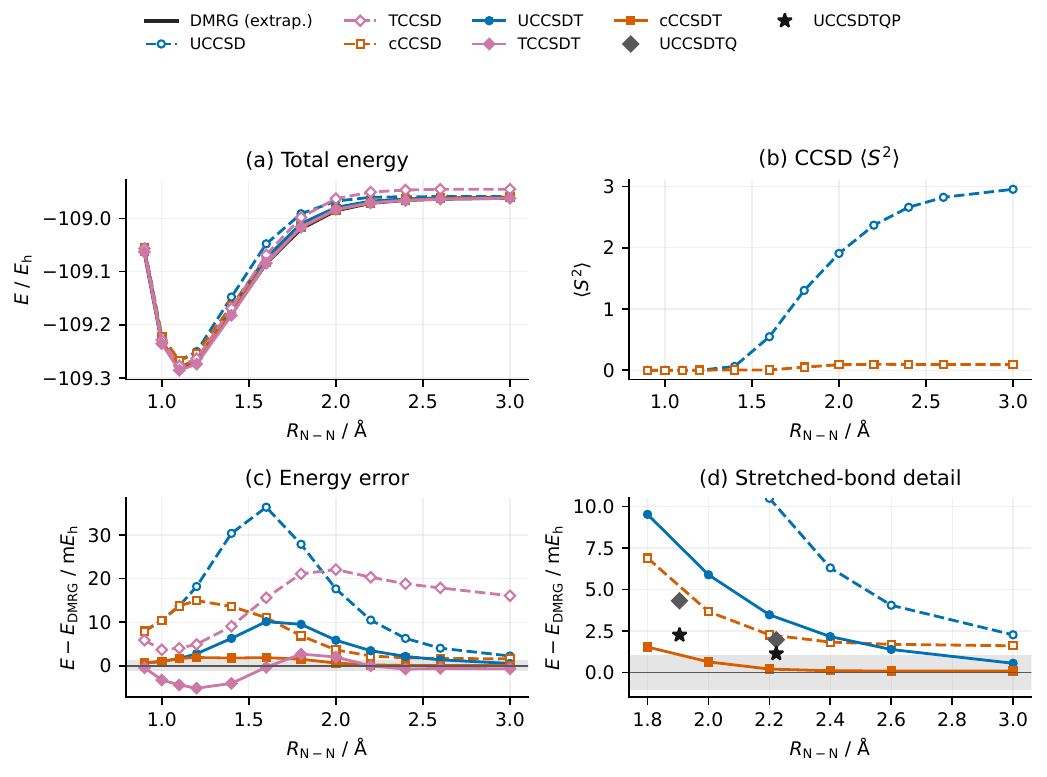}
\caption{\ce{N2} dissociation in cc-pVDZ: (a) total energies from unrestricted CC, CISD-constrained CC, and tailored CC together with the reference, extrapolated DMRG energy; (b) $\langle S^2\rangle$ from the UCCSD and cCCSD response densities; (c) energy errors of CC methods across the full bond-length grid; and (d) the stretched-bond errors for UCCSD, UCCSDT, cCCSD, cCCSDT, and literature UCCSDTQ and UCCSDTQP. 
The unrestricted and constrained calculations use UHF references, while TCCSD and TCCSDT use an RHF reference and are tailored to a spin-pure CASCI(6e,6o) wavefunction in those orbitals.
The full-grid CC error curves are referenced to extrapolated DMRG energies performed in this work. The UCCSDTQ and UCCSDTQP errors are from Table 3 by Chan et al.\cite {Chan2004N2}. The shaded regions in (c,d) mark errors within $\pm1$~$\mEh$.}
\label{fig:n2}
\end{figure}

We next turn to the standard molecular benchmark problem of bond dissociation.
We first consider \ce{N2} bond dissociation in the cc-pVDZ basis (Figure~\ref{fig:n2}). 
Near the equilibrium bond length, cCC reduces to the regular CC, since there is no symmetry breaking in the reference and the CC wavefunction is automatically spin pure. The distinction between regular UCC and cCC becomes pronounced as the bond stretches and the UHF reference breaks spin symmetry (Figure~\ref{fig:n2}a). At the dissociation limit, UCC strongly mixes high-spin states, with $\langle S^2\rangle$ approaching 3.0 (Figure~\ref{fig:n2}b). Although the spin constraint applies only in the CISD space, cCC largely removes spin contamination across all bond lengths and significantly reduces the energy error (Figure~\ref{fig:n2}b and c), with cCCSD approaching UCCSDTQ accuracy and cCCSDT beating UCCSDTQP (Figure~\ref{fig:n2}d).

As a comparison, we also consider a tailored CC (TCC) approach\cite{Kinoshita2005TailoredCC}. We freeze the active-space $t$ amplitudes obtained by cluster decomposition of a spin-pure CASCI(6e,6o) wavefunction and optimize only the external amplitudes.
Although TCCSD reduces the UCCSD error at short bond distances, it exhibits a large error at the dissociation limit. When going from TCCSD to TCCSDT, the improvement is not uniform. TCCSDT overshoots at short bond distances, making the binding curve substantially less accurate than cCCSDT. TCC also depends on the orbitals used to define the active space\cite{morchen2020tailored}, and the alternatives we have examined are less accurate (see SI).

We also use the \ce{N2} dissociation problem to examine the size consistency of cCC in the same cc-pVDZ basis. We compare the dimer energy at $R=10^{10}$~\AA\ with that from two independent high-spin nitrogen atoms using both unrestricted and restricted open-shell orbitals. Defining
\begin{align}
 \delta_{\mathrm{SC}}=E(\ce{N2};R\rightarrow\infty;S=0)-2E(\ce{N};S=3/2),
\end{align}
we obtain $\delta_{\mathrm{SC}}=-4.7\times10^{-5}$ and $-1.9\times10^{-6}~ E_{\mathrm h}$ for CISD-constrained cCCSD and cCCSDT, respectively, when using the unrestricted orbitals. We obtain $-1.1\times10^{-4}$ and $-4.5\times10^{-6}~ E_\mathrm{h}$ for restricted orbitals. Thus, with truncated CI constraints, cCC is not strictly size-consistent, but the violation is numerically small. This also shows that the small cCCSD error relative to the DMRG energy at the dissociation limit shown in Figure~\ref{fig:n2}d is due to an imperfect description of isolated nitrogen atoms at the SD level. 

\begin{figure}[htbp]
\centering
\includegraphics[width=\linewidth]{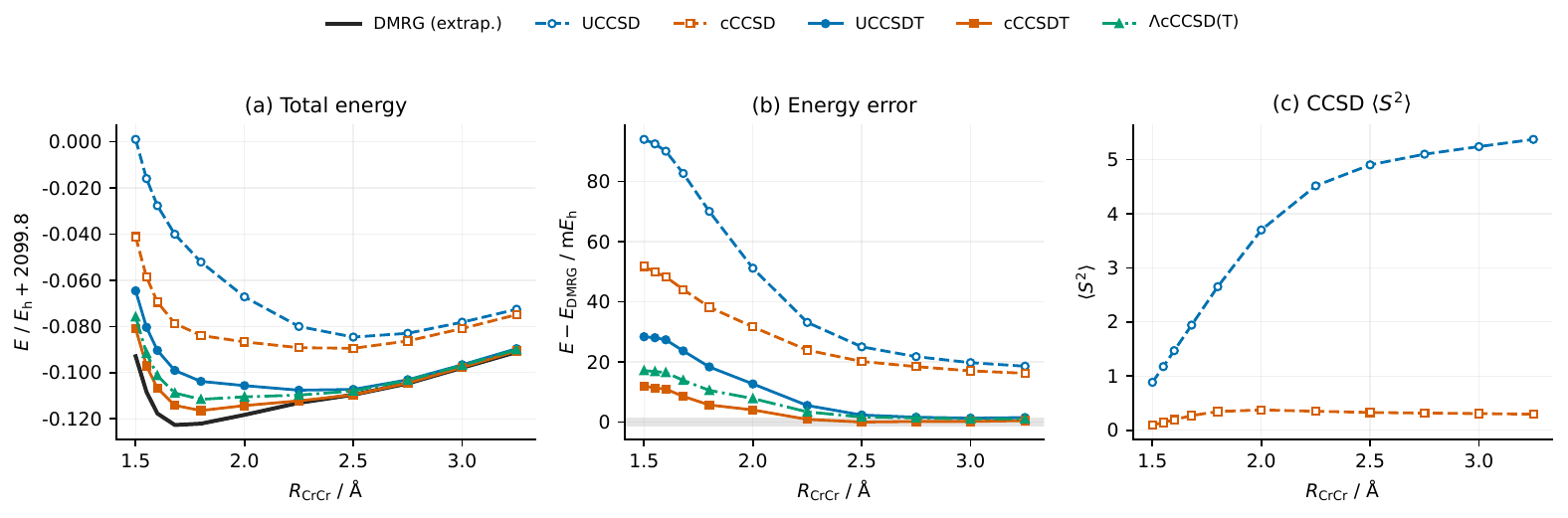}
\caption{\ce{Cr2} dissociation with X2C/cc-pVDZ-DK, correlating 28 electrons in 76 orbitals: (a) conventional and constrained CC total energies together with extrapolated DMRG; (b) CC energy errors relative to extrapolated DMRG; and (c) $\langle S^2\rangle$ evaluated from the CCSD response density. The shaded region in (b) marks errors within $\pm1$~m$E_{\mathrm h}$. The constrained methods use the CISD-space constraint. All CC calculations used the UHF reference.}
\label{fig:cr2}
\end{figure}

We next consider the chromium dimer, a demanding test.
We use the X2C Hamiltonian and cc-pVDZ-DK basis, correlating 28 electrons in 76 orbitals. Previous density matrix renormalization group (DMRG) calculations in the same setup employed very large bond dimensions (up to $D_\text{max}=28000$) and the extrapolation distance ($E(D_\text{max})-E_\text{extrap.}$) ranged from approximately 7 to 11~m$E_\text{h}$. The accuracy of the extrapolated DMRG energy was separately established from small deviations ($\le 5\times10^{-4}~E_\text{h}$) from extrapolated stochastic heat-bath CI energies\cite{larsson_chromium_2022}. We use DMRG $E_\text{extrap.}$ as the reference in Figure~\ref{fig:cr2}.


Unlike the \ce{N2} case, where UCCSD is able to reproduce the qualitative dissociation curve, unconstrained UCCSD for \ce{Cr2} is badly spin-contaminated and fails to reproduce the binding well at $\sim 1.68$~\AA. The CISD constraint largely improves both the spin and energy, reducing the energy error by up to $\sim40~\mEh$ and $\langle S^2\rangle$ by up to $\sim5$. The CISD-constrained CC at the SD level already resembles the curve shape of the unconstrained UCCSDT. The CISD-constrained CCSDT further improves the curve and reproduces the extrapolated DMRG shape. Starting at $R_\text{CrCr}=2.25$~\AA, cCCSDT agrees with extrapolated DMRG within $1~\mEh$, while at shorter bond distances, its error is comparable to that of the largest-scale DMRG variational energies\cite{larsson_chromium_2022} (at $D=18000-25000$, approximately 8--11 $\mEh$ from the extrapolated values for $R_\text{CrCr}<2.25$~\AA).




We next consider a perturbative triples approach as a cheaper alternative to the full triples. At $1.68$~\AA, the conventional (T) expression\cite{raghavachari1989fifth} evaluated with constrained amplitudes (not shown in Figure~\ref{fig:cr2}) gives a correction of $-117.0~\mEh$, making it $73~\mEh$ lower than the extrapolated DMRG energy. Because the constraint alters the CC response equations (Eq.~\ref{eq:ccc_lambda_block}), the assumption that the left response can be approximated by $T^\dagger$ appears to no longer be reliable. Instead, a $\Lambda$-CCSD(T) correction\cite{lambdat} based on the actual constrained response is more appropriate. As shown in Figure~\ref{fig:cr2}, the $\Lambda$-(T) gives a similar curve shape as cCCSDT, and only underestimates the binding energy by $4.3~\mEh$ compared to cCCSDT.

\subsection{Iron--Sulfur Clusters}

We now test the computation of spin gaps of the two-iron iron-sulfur cluster \ce{Fe2S2(SCH3)4}, in both the oxidized ferric-ferric state (total charge $=-2$) and the mixed-valence ferric-ferrous state (total charge $=-3$).
For Fe$^{3+}$Fe$^{3+}$, we consider the gap between the fully ferromagnetic $S=5$ state and the antiferromagnetic $S=0$ state, whose magnetic coupling is dominated by direct exchange and superexchange\cite{sharma2014low}. For mixed-valence Fe$^{3+}$Fe$^{2+}$, we consider the gap between $S=9/2$ and $S=1/2$, to which double exchange additionally contributes\cite{anderson1955considerations,sharma2014low}.

We first benchmark the UHF-based cCC approach in a 20-orbital active-space model\cite{li2017spin}, containing the Fe 3d orbitals, the bridging-sulfur 3p orbitals, and one 3p orbital from each terminal sulfur, where numerically exact DMRG energies are obtainable with $D_\text{max}=4000$ (all discarded weights are $<2\times10^{-11}$).

We define each gap as $\Delta E=E(S_{\mathrm{high}})-E(S_{\mathrm{low}})$. Table~\ref{tab:fe_gaps} compares the gaps against the DMRG energies.
Since the spin gaps $E(S=1)-E(S=0)$ and $E(S=3/2)-E(S=1/2)$ are quite small (on the order of a few hundred wavenumbers\cite{sharma2014low}), a spin constraint to remove spin contamination is not expected to make a huge energy correction. Despite this, the CISD constraint reduces the cCCSDT error for the Fe$^{3+}$Fe$^{3+}$ spin gap to within $1~\mEh$, whereas unconstrained CCSDT retains an error of $\sim3~\mEh$.
The smaller spin gap of the mixed-valence Fe$^{3+}$Fe$^{2+}$ cluster is consistent with additional ferromagnetic stabilization of the high-spin state through double exchange, and cCCSD, $\Lambda$cCCSD(T), and cCCSDT all achieve an accuracy within 1 kcal/mol.

Even though the energy corrections are not large, at most $3~\mEh$, of the same order as a few hundred wavenumbers, the CISD constraint significantly improves the spin contamination. The CCSD $\langle S^2\rangle$ decreases from $3.6013$ (unconstrained) to $0.1945$ (CISD-constrained) for the singlet, and from $3.2996$ to $0.8604$ for the doublet.

\begin{table}[htbp]
\centering
\caption{High-spin minus low-spin energy gaps for the 20-orbital models and full iron--sulfur problems, in m$E_{\mathrm h}$. Parentheses give signed errors relative to DMRG for the 20-orbital models. For the high-spin states in the 20-orbital models, the constraint is unnecessary as all $\alpha$ spin-orbitals are singly occupied, and thus the ordinary parent method is used. All CC calculations used the UHF reference.}
\label{tab:fe_gaps}
\begin{tabular}{lrr}
\hline
Method & Fe$^{3+}$Fe$^{3+}$: $S=5$ vs.\ $0$ & Fe$^{3+}$Fe$^{2+}$: $S=9/2$ vs.\ $1/2$ \\
\hline
\multicolumn{3}{l}{20-orbital model} \\
UCCSD & 19.189 ($-8.886$) & 17.192 ($-2.744$) \\
UCCSDT & 25.315 ($-2.760$) & 19.042 ($-0.894$) \\
cCCSD & 21.804 ($-6.271$) & 19.143 ($-0.793$) \\
$\Lambda$cCCSD(T) & 25.464 ($-2.611$) & 18.578 ($-1.358$) \\
cCCSDT & 27.539 ($-0.536$) & 20.570 ($+0.634$) \\
DMRG & 28.075 & 19.936 \\
\hline
\multicolumn{3}{l}{Full problem} \\
cCCSD & 22.472 & 7.909 \\
$\Lambda$cCCSD(T) & 29.017 & 8.215 \\
\hline
\end{tabular}
\end{table}

\begin{figure}[htbp]
\centering
\includegraphics[width=\linewidth]{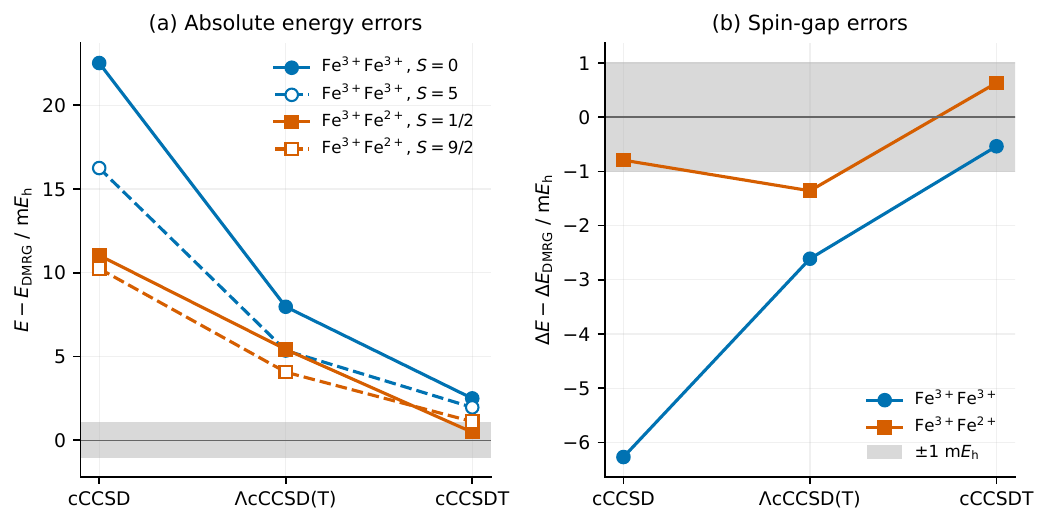}
\caption{Errors relative to DMRG for the 20-orbital iron--sulfur models: (a) absolute energies of the low- and high-spin states and (b) high-spin minus low-spin gaps. The low-spin and high-spin states correspond to $S=0,5$ for Fe$^{3+}$Fe$^{3+}$ and $S=1/2,9/2$ for Fe$^{3+}$Fe$^{2+}$. The shaded regions mark errors within $\pm1$~m$E_{\mathrm h}$. Lines are used to guide the eye between the three methods.}
\label{fig:fe_benchmarks}
\end{figure}

Figure~\ref{fig:fe_benchmarks} separates absolute-energy errors from spin-gap errors. The constrained absolute energies improve systematically from doubles through $\Lambda$-based perturbative triples to full triples. For the Fe$^{3+}$Fe$^{2+}$ doublet, a sub~m$E_{\mathrm h}$ accuracy in absolute energy is achieved by cCCSDT.

We finally move beyond the active-space benchmark and compute the spin gaps in the full orbital problem.
We employ the TZP-DKH basis for the irons and the bridging sulfurs and DZP-DKH for other atoms, and an X2C Hamiltonian on the geometry taken from Supplementary Table 1 of Ref.~\cite{sharma2014low} with 44 frozen core orbitals computed from CASSCF. We perform cCC calculations on UHF references. The full iron--sulfur calculations ran on a two-socket Intel Ice Lake node using 64 physical cores and 64 OpenMP threads. Across the four spin states (128 cCCSD iterations in total), the nonconstraint portion of an iteration took 780~s on average, while the CISD-constraint evaluation and derivatives took 36~s. The constraint work therefore added 4.6\% relative to the regular CCSD iteration work and accounted for 4.4\% of the total cCCSD iteration wall time.

In terms of spin purity, the cCCSD response $\langle S^2\rangle$ values indicate almost pure spins for the high-spin states: $\langle S^2\rangle=30.005$ and $2S_\mathrm{eff}+1=11.001$ for the oxidized $S=5$ state, and $\langle S^2\rangle=24.756$ and $2S_\mathrm{eff}+1=10.001$ for the mixed-valence $S=9/2$ state. The low-spin states are less spin-pure: $\langle S^2\rangle=0.489$ and $2S_\mathrm{eff}+1=1.720$ for the oxidized singlet, and $\langle S^2\rangle=1.098$ and $2S_\mathrm{eff}+1=2.322$ for the doublet. Nevertheless, these effective multiplicities are sufficient for the spin-state assignments to be unambiguous.

For Fe$^{3+}$Fe$^{3+}$, the cCCSD gap changes by only $0.668~\mEh$ between the 20-orbital model and the full calculation. Assuming that the higher-level correlation correction obtained in the 20-orbital model is transferable to the full orbital space, we form the composite estimate $\Delta E_\text{cCCSD}(\text{full})-\Delta E_\text{cCCSD}(\text{20o})+\Delta E_\text{DMRG}(\text{20o})=28.743~\mEh$. This estimate closely agrees with the full-space $\Lambda$cCCSD(T) value of $29.017~\mEh$.
For Fe$^{3+}$Fe$^{2+}$, the $\Lambda$cCCSD(T) gap agrees with DMRG within 1 kcal/mol in the 20-orbital benchmark. Assuming that a similar accuracy carries over to the full orbital space, we take the full-space $\Lambda$cCCSD(T) result as our best estimate. Similar to the 20-orbital model, the full-space cCCSD and $\Lambda$cCCSD(T) gaps differ by only $0.306~\mEh$ ($7.909~\mEh$ vs. $8.215~\mEh$), indicating that the net perturbative-triples correction to this gap is small.
We therefore estimate the full-problem Fe$^{3+}$Fe$^{3+}$ and Fe$^{3+}$Fe$^{2+}$ spin gaps to be approximately $29$ and $8~\mEh$, respectively. Relative to the 20-orbital results, enlargement of the correlated orbital space has little effect on the oxidized-state gap but substantially reduces the mixed-valence gap. Using the Heisenberg spin-ladder expression, $E(S)=JS(S+1)$, the $29~\mEh$ Fe$^{3+}$Fe$^{3+}$ gap defines an effective coupling $J_\mathrm{eff}\approx210$ cm$^{-1}$. This is comparable to the spin-ladder fit $J\approx236$ cm$^{-1}$ from DMRG-SCF\cite{sharma2014low} and the experimental fit $J\approx148$ cm$^{-1}$ for a related synthetic dimer.\cite{gillum1976synthetic} The $8~\mEh$ Fe$^{3+}$Fe$^{2+}$ gap corresponds to approximately $1800$ cm$^{-1}$, compared to approximately $1100$ cm$^{-1}$ by DMRG-SCF with a 32-orbital active space\cite{sharma2014low}.
Taken together, these calculations move beyond small active-space models and provide direct estimates of chemically relevant total spin gaps in the full orbital problem, showing that the mixed-valence spin gap, with its double-exchange contribution, is sensitive to the correlated ligand and iron orbital space.


\section{Conclusions}

Strongly correlated electrons have long posed a central challenge for single-reference coupled cluster theory. In this work, we address this challenge by supplementing the usual CC equations with physically motivated numerical constraints. To this end, we developed a general constrained CC framework that allows arbitrary constraints to be imposed on the CC solution without changing the underlying CC diagrammatic algebra.

We demonstrate that cCC equipped with a CISD-space spin-symmetry constraint significantly improves conventional CC energies and spin purity across a wide range of strongly correlated problems. The construction applies at arbitrary CC truncation levels, accommodates unrestricted symmetry-broken references and many-open-shell low-spin states, and adds constraint contractions with only $\mathcal O(N^4)$ computational cost.

The present results establish spin-constrained CC as a promising first realization of this broader framework. More generally, cCC provides a flexible foundation for developing new coupled cluster methods that encode physical information beyond spin to address other manifestations of strong correlation.

\section*{Code availability}

The constrained coupled cluster implementation used in this work is available at \url{https://github.com/MoleOrbitalHybridAnalyst/constrained-coupled-cluster}.

\section*{Supporting information}

The following files are available free of charge.
\begin{itemize}
  \item \texttt{si.pdf}: Computational details and convergence diagnostics for the Hubbard, nitrogen, chromium, and iron--sulfur calculations; numerical solution of the constrained coupled-cluster amplitude and response equations; and comparisons of tailored CC using different orbital choices.
  \item \texttt{si\_data.zip}: Numerical data underlying all figures in the main text and Supporting Information, in CSV format, together with a Python script that regenerates the figures.
\end{itemize}

\printbibliography

\newpage

\rule{0.05in}{1.75in}%
\begin{minipage}[b][1.75in]{3.25in}
  \sffamily
  \frenchspacing

  Some journals require a graphical entry for the Table of Contents. This
  should be laid out ``print ready'' so that the sizing of the text is correct.

  The space available depends on the journal: J. Am. Chem. Soc. allows 3.25 in
  by 1.75 in and requires sanserif text. Some journals want different sizes:
  you can easily adjust here.

  The two rules either side of the content are there to help judge the height
  of your material: they may be deleted once not required.

\end{minipage}%
\rule{0.05in}{1.75in}

\end{document}


\maketitle
\begingroup\renewcommand\thefootnote{$\dagger$}\footnotetext{Current address: Department of Molecular Biology and Biochemistry, University of California, Irvine, California 92697, United States}\endgroup

\section{DIIS-based Solution of the Constrained Amplitude Equations}

We describe the numerical procedures used to solve the constrained amplitude and response equations. Let $t$ denote the independent cluster amplitudes, $R(t)$ the conventional coupled-cluster (CC) residual vector, and $C(t)$ a vector of $m$ constraints. The constraint Jacobian is
\begin{align}
 G_{A\mu}(t)=\frac{\pd C_A(t)}{\pd t_\mu}.
\end{align}
The coupled amplitude and multiplier equations are
\begin{align}
 F(t,\eta)
 =\begin{pmatrix}
 R(t)-G(t)^{\mathrm T}\eta\\
 C(t)
 \end{pmatrix}
 =0.
 \label{eq:si_coupled_residual}
\end{align}
Here, $\eta$ contains one multiplier for each constraint. Only independent amplitudes that enter the constraints are included in the constraint operations, with the appropriate fermionic permutation signs retained when translating between these coordinates and the full amplitude tensors.

\subsection{Constraint-Space Correction to a Conventional CC Iteration}

A practical solution of the cCC equations begins from the ordinary diagonally preconditioned CC update $\widetilde t$. Let $D_\mu$ be the diagonal approximation to the CC Jacobian, i.e., the orbital energy gap. Linearization of Eq.~\ref{eq:si_coupled_residual} about the current amplitudes gives the following update rules
\begin{align}
 t^{\mathrm{new}}&=\widetilde t+D^{-1}G^{\mathrm T}\eta,\label{eq:si_update}\\
 (GD^{-1}G^{\mathrm T})\eta
 &=-C-G(\widetilde t-t).\label{eq:si_gram}
\end{align}
For notational simplicity, we define
\begin{align}
 M=GD^{-1}G^{\mathrm T}    \\
 b=-C-G(\widetilde t-t)
\end{align}

We used complementary dense and matrix-free implementations of the constraint correction. The matrix-free route evaluates Jacobian-vector products (JVPs) and vector-Jacobian products (VJPs) without storing $G$ or $M$, thereby retaining the formal scaling of the underlying contractions. The dense route explicitly forms these matrices and was used for sufficiently small constraint spaces.

In the matrix-free implementation, Eq.~\ref{eq:si_gram} is solved by preconditioned conjugate gradients. A Jacobi preconditioner is constructed from the diagonal of $M$, and the multiplier vector from the preceding CC iteration is used as the initial guess when available. In the dense implementation, the symmetric eigendecomposition of $M$ provides a direct solution. This decomposition is particularly convenient when a step restriction is needed, because trial shifts can be tested by changing only the eigenvalue denominators.

Specifically, when the constraint-induced amplitude correction in Eq.~\ref{eq:si_update} exceeds the maximum allowed value $\Delta$, we introduce a Levenberg--Marquardt shift,
\begin{align}
 (M+\mu I)\eta=b.
 \label{eq:si_shifted_gram}
\end{align}
The smallest nonnegative shift satisfying $\max(|D^{-1}G^\mathrm{T}\eta|)\leq\Delta$ is located by bracketing and bisection. The shift damps poorly conditioned directions that can arise from small singular values of $G$, especially for CSF constraints. In the dense routine, every trial value of $\mu$ is evaluated using the same eigendecomposition of $M$ while the matrix-free routine instead performs shifted conjugate-gradient solves.

For highly nonlinear constraints, the solution to Eq.~\ref{eq:si_gram} that enforces $C(t)=0$ to the first order may be refined by minimizing the nonlinear constraint residual while holding $\widetilde t$ and $G$ fixed,
\begin{align}
 \phi(\eta)
 =\frac{1}{2}\left\|C(\widetilde t+D^{-1}G^\mathrm{T}\eta)\right\|^2.
 \label{eq:si_nonlinear_eta}
\end{align}
Again, the minimization is optionally subject to the trust-region bound $\max(|D^{-1}G^\mathrm{T}\eta|)\leq\Delta$. In the dense routine, we use sequential least-squares programming with linear amplitude-step bounds. The matrix-free treatment employs Gauss--Newton steps, a least-squares minimal-residual (LSMR) solver, and backtracking.

The resulting iterative procedure is summarized below.
\begin{enumerate}
 \item Evaluate the constraint residual $C(t)$ and the conventional CC update $\widetilde t$.
 \item Construct the actions of $G$, $G^{\mathrm T}$, and $M=GD^{-1}G^{\mathrm T}$.
 \item Solve Eq.~\ref{eq:si_gram}, introducing the smallest trust-region shift required by Eq.~\ref{eq:si_shifted_gram}.
 \item When used, refine the multipliers by minimizing Eq.~\ref{eq:si_nonlinear_eta} in the constraint space.
 \item Form $t\leftarrow\widetilde t+D^{-1}G^{\mathrm T}\eta$, reevaluate $C(t)$, and apply DIIS.
 \item Repeat until the energy, amplitude-update norm, and constraint residual meet their respective convergence thresholds.
\end{enumerate}

\subsection{DIIS and Convergence Criteria}

We implemented two flavors of constraint-augmented DIIS.

\subsubsection{Augmented DIIS Error Vector}
In the first option, the DIIS error vector combines the ordinary amplitude-update error with row-scaled constraint residuals,
\begin{align}
 e_{A}=w_C\frac{C_A}{\|G_{A:}\|}.
 \label{eq:si_diis}
\end{align}
where $w_C$ is a tunable constraint DIIS weight. Such a row normalization prevents the arbitrary scale of an individual constraint from dominating the DIIS extrapolation.
Constraint rows with near-zero Jacobian norm were omitted from this augmented error vector.

\subsubsection{Tangent-Space DIIS}
The augmented error vector asks DIIS to reduce the CC and constraint residuals simultaneously. For difficult calculations, we instead use tangent-space DIIS to separate these two tasks. The explicit correction in Eq.~\ref{eq:si_update} restores the constraints, while tangent-space DIIS accelerates the remaining motion along the constraint manifold.

This separation follows directly from the linearized constraints. For an amplitude-space direction $e$,
\begin{align}
 C(t+\epsilon e)
 =C(t)+\epsilon Ge+\mathcal O(\epsilon^2).
\end{align}
Consequently, $Ge=0$ means that motion along $e$ does not change the constraints to first order. We thus want $Ge=0$, i.e., a DIIS error vector $e$ to live in the constraint tangent space.

Starting from the conventional CC update $\widetilde t-t$, we seek a tangent error of the form
\begin{align}
 e^{\mathrm{tan}}
 =\widetilde t-t+D^{-1}G^{\mathrm T}\eta^{\mathrm{tan}}.
 \label{eq:si_tangent_diis_error}
\end{align}
Imposing $Ge^{\mathrm{tan}}=0$ gives
\begin{align}
 \eta^{\mathrm{tan}}
 =-\left(GD^{-1}G^{\mathrm T}\right)^{+}
 G(\widetilde t-t).
 \label{eq:si_tangent_diis_multiplier}
\end{align}
Thus, $\eta^{\mathrm{tan}}$ is an auxiliary quantity chosen to remove the constraint-changing component of the $t$ update, instead of minimizing the cCC residual. When $GD^{-1}G^{\mathrm T}$ is rank deficient, the pseudoinverse gives the minimum-norm solution that minimizes the remaining violation of the tangency condition. Amplitudes absent from the constraints retain their ordinary CC update.

At cCC iteration $i$, the pre-DIIS $t$ vector
\begin{align}
 t_i^{\mathrm{corr}}
 =\widetilde t_i+D_i^{-1}G_i^{\mathrm T}\eta_i
\end{align}
still uses $\eta_i$ from the constraint correction in Eq.~\ref{eq:si_gram} or its nonlinear refinement, not the auxiliary $\eta_i^{\mathrm{tan}}$. The standard Pulay coefficients are obtained from
\begin{align}
 \mathcal D_{ij}
 &=(e_i^{\mathrm{tan}})^{\mathrm T}e_j^{\mathrm{tan}},\nonumber\\
 t^{\mathrm{DIIS}}
 &=\sum_{i=1}^{n}c_i t_i^{\mathrm{corr}},\nonumber\\
 \sum_{i=1}^{n}c_i&=1,
 \label{eq:si_tangent_diis_pulay}
\end{align}
with the coefficients $c_i$ chosen to minimize the norm of the extrapolated tangent error.

All the calculations used the augmented error vector in Eq.~\ref{eq:si_diis} unless the tangent-space DIIS is specified.

\subsection{cCC Convergence}
Convergence of cCC is accepted only when the final amplitudes meet both the parent energy and amplitude criteria and $\max(|C(t)|)\leq\epsilon_C$. The constraint residual is reevaluated after the final DIIS extrapolation, and one additional constraint update without DIIS is optionally performed. The calculation is considered unconverged if this update fails to reach the constraint threshold or changes the energy or $t$ amplitude norm by more than ten times the corresponding parent threshold.

\section{Joint-Newton Solver for cCC}
\label{sec:si_joint_newton}

The constraint-space iteration is sufficient for most calculations, but it can stall for strongly nonlinear constraints because the CC update and the constraint correction are determined separately. We therefore use a joint-Newton solver as a final refinement when necessary. The idea is to update the amplitudes $t$ and multipliers $\eta$ together so that the constrained amplitude residual and the constraint residual decrease in the same step.

The effective amplitude Jacobian is
\begin{align}
 K_{\mu\nu}(t,\eta)
 &=
 \frac{\pd R_\mu(t)}{\pd t_\nu}
 -
 \sum_A\eta_A
 \frac{\pd^2 C_A(t)}
 {\pd t_\mu\pd t_\nu}.
\end{align}
The joint Newton step then satisfies
\begin{align}
 \begin{pmatrix}
 K & -G^{\mathrm T}\\
 G & 0
 \end{pmatrix}
 \begin{pmatrix}
 \delta t\\
 \delta\eta
 \end{pmatrix}
 =-
 \begin{pmatrix}
 R-G^{\mathrm T}\eta\\
 C
 \end{pmatrix}.
 \label{eq:si_newton}
\end{align}
The first row improves the constrained CC amplitude equations, while the second restores the constraints. The off-diagonal blocks express their mutual dependence: $G\delta t$ is the first-order change in the constraints, and $-G^{\mathrm T}\delta\eta$ is the change in the amplitude residual produced by the multipliers. The constraint-Hessian contribution in $K$ accounts for the curvature of a nonlinear constraint manifold.

For small problems, we form this coupled matrix explicitly and use a rank-revealing least-squares solution, which remains well-defined when some constraint directions are redundant. For larger problems, we apply the same matrix only through matrix-vector products and solve the step with damped LSMR. 

\section{Numerical Solution of the Constrained Response Equations}

Rather than solving the block response system in the main text directly, we eliminate $\kappa$ and solve for $\lambda$ in the tangent space of the constraints. A tangent-space projector is
\begin{align}
 \Pi_{\mathrm{tan}}
 =I-G^{\mathrm T}(GG^{\mathrm T})^{-1}G.
 \label{eq:si_projector}
\end{align}
We impose $\lambda=\Pi_{\mathrm{tan}}\lambda$ and solve
\begin{align}
 \Pi_{\mathrm{tan}}
 \left(K^{\mathrm T}\lambda+e\right)
 =0
 \label{eq:si_projected_lambda}
\end{align}
with preconditioned LGMRES.

After $\lambda$ converges, we define the unprojected residual
\begin{align}
 \rho=K^{\mathrm T}\lambda+e
\end{align}
and recover the eliminated multiplier from
\begin{align}
 \kappa=-(GG^{\mathrm T})^{-1}G\rho.
 \label{eq:si_kappa}
\end{align}
Systems involving $GG^{\mathrm T}$ are solved directly when the constraint matrices are stored and by Jacobi-preconditioned conjugate gradients in the matrix-free implementation.

\section{Computational Details}

All self-consistent-field (SCF), exact-diagonalization, and coupled-cluster calculations reported in the main text were performed with PySCF and our own PySCF-based implementations. The constrained coupled-cluster implementation is available at \url{https://github.com/MoleOrbitalHybridAnalyst/constrained-coupled-cluster}; the tailored coupled-cluster calculations used a separate in-house implementation. The 20-orbital density matrix renormalization group (DMRG) calculations employed Block2. Throughout this section, spin labels follow the PySCF convention, \texttt{spin} $=N_\alpha-N_\beta$. Unless noted otherwise, cCCSD and cCCSDT denote unrestricted calculations employing the CISD-space spin constraint, and TCCSD and TCCSDT denote tailored calculations using an RHF reference and a spin-pure CASCI in those orbitals. CCSDT calculations were initialized from the converged amplitudes of the corresponding CCSD calculation, with the $T_3$ amplitudes initially set to zero. The reported biorthogonal CCSD values of $\langle S^2\rangle$ were evaluated from the one- and two-particle response density matrices after solution of the appropriate $\Lambda$ equations. Because a CCSDT response density was not evaluated, corresponding response values of $\langle S^2\rangle$ are not reported.

\subsection{Hubbard Models}

The Hubbard benchmarks comprised an eight-site open chain and an open $2\times4$ lattice. The one-dimensional model used nearest-neighbor hopping $t=1$ and on-site repulsion $U$; in the two-dimensional lattice, sites were ordered row by row and horizontal and vertical nearest-neighbor matrix elements were set to $-1$. Each system contained eight electrons, with four $\alpha$ and four $\beta$ electrons. We examined the interaction strengths
\begin{align}
 U/t=1,2,4,6,8,10,12,16,20.
\end{align}
At each interaction strength, UHF was initialized from a checkerboard N\'eel density and converged to an energy threshold of $10^{-12}~t$. Both lattices were studied with UCCSD, UCCSDT, cCCSD, and cCCSDT. For the one-dimensional chain, we additionally performed a matrix-free CISDT-cCCSDT calculation.

The CC convergence thresholds were $10^{-8}~t$ for the energy and $10^{-5}$ for the amplitude-update norm. Direct inversion in the iterative subspace (DIIS) was initiated after five iterations with a maximum subspace dimension of 20, and a level shift of $10^{-3}~t$ was applied. Constrained calculations used a constraint threshold of $10^{-7}$ and the nonlinear multiplier solution of Eq.~\ref{eq:si_nonlinear_eta}. The maximum allowed constraint-induced amplitude update was 0.1. Exact reference energies were obtained by diagonalizing the complete 4900-determinant Hamiltonian in the $M_S=0$ sector.

The fixed-reference comparison used the same eight-site open chain at $U/t=16$ and the site-orbital N\'eel determinant $|\alpha\beta\alpha\beta\alpha\beta\alpha\beta\rangle$. The CC settings were unchanged. Constraints were constructed from the CISD and CISDT spaces associated with this reference or from the target singlet configurational state function (CSF). Additional settings for the joint-Newton refinement of the CSF-constrained CCSD result are provided in Section~\ref{sec:si_hubbard_csf}.

Ket-state diagnostics were obtained by explicitly applying the exponential $\exp(T)$ to the reference determinant. The normalized ket was used for the full-space fidelity. For the CISD-space fidelity, both the CC and exact states were projected into the reference-plus-singles-and-doubles space and renormalized before their overlap was evaluated.

\subsubsection{Convergence of the Eight-Site CSF-Constrained Calculation}
\label{sec:si_hubbard_csf}

The CSF-constrained CCSD calculation for the eight-site, eight-electron open chain at $U/t=16$ required additional refinement with the joint-Newton procedure of Eq.~\ref{eq:si_newton}. The calculation used the fixed site-orbital N\'eel determinant. The CSF constraint comprised 56 equations acting on 360 independent UCCSD amplitudes.

An initial constraint-space iteration used an energy threshold of $10^{-8}\,t$, an amplitude-update norm threshold of $10^{-5}$, and a constraint threshold of $10^{-7}$. DIIS was initiated after five iterations with a maximum dimension of 20, and a level shift of $10^{-3}\,t$ was applied. The maximum allowed constraint-induced amplitude update was 0.1, and the nonlinear multiplier solution of Eq.~\ref{eq:si_nonlinear_eta} was employed. This calculation reduced the maximum constraint residual to $8.23\times10^{-11}$. However, the joint amplitude residual remained $7.29\times10^{-5}$ and did not satisfy Eq.~\ref{eq:si_coupled_residual} to the desired accuracy.

We therefore refitted the multipliers by least squares and refined the solution with the dense joint-Newton procedure. Table~\ref{tab:si_hubbard_newton} summarizes the numerical settings and final diagnostics. The coupled matrix had dimension 416 and numerical rank 414 at the stated cutoff, necessitating a rank-revealing least-squares solution. Eleven accepted Newton steps reduced both residual blocks below their target tolerances. This jointly converged solution provided the energy, spin, and fidelity values reported in the main text.

\begin{table}[htbp]
\centering
\caption{Joint-Newton settings and final diagnostics for the CSF-constrained CCSD calculation at $U/t=16$.}
\label{tab:si_hubbard_newton}
\begin{tabular}{lc}
\toprule
Quantity & Value\\
\midrule
Linear algebra method & dense rank-revealing least squares\\
Relative singular-value cutoff & $10^{-13}$\\
Maximum amplitude step, $\max(|\delta t|)$ & 0.1\\
Maximum trial steps per iteration & 8\\
Target $\epsilon_R$ & $10^{-10}$\\
Target $\epsilon_C$ & $10^{-11}$\\
Coupled dimension / numerical rank & 416 / 414\\
Accepted Newton steps & 11\\
Residual evaluations & 31\\
Final $\|R-G^{\mathrm T}\eta\|_\infty$ & $1.94\times10^{-13}$\\
Final $\|C\|_\infty$ & $2.52\times10^{-12}$\\
Final cCC energy & $-1.3597962381269\,t$\\
\bottomrule
\end{tabular}
\end{table}

\subsection{Nitrogen Dissociation}

The \ce{N2} dissociation curve was evaluated in the cc-pVDZ basis at the bond lengths
\begin{align}
 R/\text{\AA}=0.9,1.0,1.1,1.2,1.4,1.6,1.8,2.0,2.2,2.4,2.6,3.0.
\end{align}
All 14 electrons in 28 orbitals were correlated, with no frozen orbitals. UCCSD, UCCSDT, cCCSD, and cCCSDT used internally stable UHF references. TCCSD and TCCSDT used RHF references converged to $10^{-11}\,E_{\mathrm h}$ with the second-order solver. The CC energy and amplitude-update thresholds were $5\times10^{-8}\,E_{\mathrm h}$ and $5\times10^{-6}$, respectively, and the DIIS dimension was 30. The cCC calculations used 147 CISD spin constraints, a constraint threshold of $10^{-7}$, and the linear multiplier solution. The maximum allowed constraint-induced amplitude update was 0.1. All six amplitude calculations converged at every geometry.

The tailored calculations used the lowest singlet CASCI(6,6) state in the RHF orbitals, with the six active orbitals following four inactive occupied orbitals. Twenty roots were requested, and the lowest singlet root was selected explicitly. The CASCI wavefunction was decomposed into active-space cluster amplitudes. The active-space $T_1$ and $T_2$ amplitudes were fixed in TCCSD, while TCCSDT additionally fixed the active-space $T_3$ amplitudes. These blocks were excluded from the iterative updates.

The singlet reference curve was computed with spin-adapted DMRG using \texttt{SU(2)} symmetry in meta-L\"owdin orthogonalized atomic orbitals with Fiedler ordering. All 14 electrons and 28 orbitals were retained. The maximum bond dimension was $D_\text{max}=2000$ at every geometry except $R=0.9$~\AA, for which $D_\text{max}=2500$ was used. The reported reference energies were obtained by linear extrapolation to zero discarded weight from the last sweeps at three descending bond dimensions, $0.9\times D_\text{max}$, $0.8\times D_\text{max}$, and $0.7\times D_\text{max}$, with zero noise. Every extrapolation distance, $E(D_\text{max})-E(\text{extrap.})$, was below $0.1\,\mathrm{m}E_{\mathrm h}$.

The CCSDTQ and CCSDTQP comparison points were taken from the published cc-pVDZ correlation energies at $R=3.6$ and 4.2 Bohr. Their errors were formed by direct subtraction of the DMRG($M=4000$) correlation energies tabulated at the same geometries.

Size consistency was examined in the cc-pVDZ basis by comparing the singlet dimer at an N--N separation of $10^{10}$~\AA\ with two independently calculated high-spin nitrogen atoms. The atomic and molecular calculations used $\texttt{spin}=3$ and $\texttt{spin}=0$, respectively. In the unrestricted orbital test, after convergence of the atomic UHF solution, the molecular initial guess was formed by placing the $\alpha$-rich atomic density on one center and its spin-interchanged partner on the other. In the restricted orbital test, we converged the ROHF for individual atoms, then formed the \ce{N2} mean-field reference using the direct product of the two atomic ROHF wavefunctions with opposite spin, i.e., $|\text{N}_1,M_S=3/2\rangle|\text{N}_2,M_S=-3/2\rangle$. The UHF/ROHF energy threshold was $10^{-12}\,E_{\mathrm h}$. UCCSD, UCCSDT, cCCSD, and cCCSDT calculations on both references were performed independently for the atom and dimer. The corresponding CC energy and amplitude-update thresholds were $10^{-10}\,E_{\mathrm h}$ and $10^{-7}$. Constrained calculations used a threshold of $10^{-10}$ and the nonlinear multiplier solution. The maximum allowed constraint-induced amplitude update was 0.1.

\subsection{Chromium Dimer}

The \ce{Cr2} dissociation curve was computed with the spin-free X2C Hamiltonian and the cc-pVDZ-DK basis at the bond lengths
\begin{align}
 R/\text{\AA}=1.500,1.550,1.600,1.680,1.800,2.000,2.250,2.500,
 2.750,3.000,3.250.
\end{align}
For $R=1.500$--$2.750$~\AA, singlet CASSCF(12,12) calculations were initialized from RHF orbitals converged to $10^{-12}\,E_{\mathrm h}$. At 3.000 and 3.250~\AA, Pipek--Mezey-localized UCCSD unrestricted natural orbitals provided the initial guess. This choice maintained a continuous CASSCF solution along the dissociation coordinate. The CASSCF energy threshold was $10^{-8}\,E_{\mathrm h}$, and an additional spin penalty selected the $S=0$ state.

The ten lowest-energy CASSCF orbitals, corresponding to the Cr 1s, 2s, and 2p semicore shells on the two centers, were incorporated into a frozen-core effective Hamiltonian. The resulting correlated problem contained 28 electrons in 76 orthonormal orbitals. UHF and all subsequent CC calculations were performed directly with this Hamiltonian, without freezing additional orbitals. To identify the lowest stable broken-symmetry reference, we independently converged five UHF initial guesses: CASSCF-based antiferromagnetic and closed-shell densities, superposed spin-opposed atomic Cr densities, the RHF density, and the density obtained from the external instability of RHF. Each candidate was converged to $10^{-11}\,E_{\mathrm h}$, and the lowest-energy internally stable solution was retained.

For UCCSD and cCCSD, the energy and amplitude-update thresholds were $10^{-7}\,E_{\mathrm h}$ and $10^{-5}$. DIIS was initiated after five iterations with a maximum dimension of 40, and a level shift of $10^{-3}\,E_{\mathrm h}$ was applied. The constrained calculations used a threshold of $10^{-6}$ and the nonlinear multiplier solution. UCCSDT and cCCSDT used energy and amplitude-update thresholds of $10^{-7}\,E_{\mathrm h}$ and $5\times10^{-6}$, respectively, with a maximum DIIS dimension of 20 beginning after five iterations and the same level shift. The $T_1$ and $T_2$ amplitudes were initialized from the corresponding converged UCCSD calculation, and $T_3$ was initialized to zero. cCCSDT employed the linear multiplier solution. The maximum allowed constraint-induced amplitude update was 0.1 for both constrained methods.

\subsection{Iron--Sulfur Clusters}

We first considered the 20-orbital \ce{Fe2S2(SCH3)4} model available at \url{https://github.com/zhendongli2008/Active-space-model-for-Iron-Sulfur-Clusters/}. The oxidized and mixed-valence models contained 30 and 31 electrons, respectively, with low-/high-spin pairs $\texttt{spin}=0/10$ and $\texttt{spin}=1/9$. High-spin UHF references were converged directly with a second-order solver. Each antiferromagnetic low-spin initial guess was constructed by spin-flipping the singly occupied $d$ orbitals of one of the irons. The resulting spin-flipped densities were subsequently converged with UHF. An energy threshold of $10^{-11}\,E_{\mathrm h}$ was used throughout.

For the 20-orbital UCCSD, UCCSDT, cCCSD, and cCCSDT calculations, the energy and amplitude-update thresholds were $10^{-9}\,E_{\mathrm h}$ and $2\times10^{-7}$. DIIS was initiated after three iterations with a maximum dimension of 20, and a level shift of $10^{-3}\,E_{\mathrm h}$ was applied. Constrained calculations used a threshold of $10^{-8}$. The maximum allowed constraint-induced amplitude update was 0.1. The DMRG benchmarks used $\texttt{SU(2)}$ symmetry, $D_\text{max}=4000$, no more than 40 sweeps, and the default Block2 sweep schedule.

The full-orbital iron--sulfur calculations employed the spin-free X2C Hamiltonian and density fitting, with TZP-DKH basis sets on both Fe atoms and the two bridging sulfur atoms and DZP-DKH basis sets on all remaining atoms, for a total of 426 atomic orbitals. For each charge state, a common spatial-orbital basis for the two target spins was optimized by equal-weight state-averaged CASSCF. We used CASSCF(10,10) over the $S=5$ and $S=0$ states for charge $-2$, and CASSCF(11,10) over the $S=9/2$ and $S=1/2$ states for charge $-3$. High-spin UCCSD unrestricted natural orbitals furnished the initial orbital guesses.

The parent $\texttt{spin}=10$ UHF determinant was converged from the atomic initial guess generated by PySCF. To construct the $\texttt{spin}=9$ reference, the ten singly occupied orbitals of this determinant were localized, and ten trial densities were independently converged by placing the additional $\beta$ electron in each localized direction. The lowest-energy solution was retained. The $\texttt{spin}=0$ and $\texttt{spin}=1$ determinants were then generated by reversing the appropriate contribution on the second Fe center in the respective $\texttt{spin}=10$ and $\texttt{spin}=9$ parent states. These atomic-orbital-basis UHF calculations employed density fitting, the second-order solver, and an energy threshold of $10^{-9}\,E_{\mathrm h}$.

Following the state-averaged optimization, the 44 lowest-energy core orbitals were incorporated into the one-electron Hamiltonian. Each selected orbital was verified to have a meta-L\"owdin weight of at least 0.99 in the C 1s, S 1s/2s/2p, or Fe 1s/2s/2p shells. The resulting correlated Hamiltonians contained 382 orbitals and 98 electrons for charge $-2$, or 99 electrons for charge $-3$. Even-tempered Gaussian auxiliary basis sets were used for both UHF and the orbital transformations.

The spin-specific UHF determinants were projected into the common orbital basis, converted to idempotent densities with the required numbers of $\alpha$ and $\beta$ electrons, and reconverged in the 382-orbital effective Hamiltonian with the second-order UHF solver. The UHF energy threshold was $10^{-12}\,E_{\mathrm h}$. The reported full-orbital results were obtained with density-fitted cCCSD and $\Lambda$cCCSD(T). For CCSD, the energy and amplitude-update thresholds were $10^{-7}\,E_{\mathrm h}$ and $5\times10^{-6}$. DIIS was initiated after three iterations with a maximum dimension of 20. The constrained calculations used a threshold of $10^{-7}$ and the linear multiplier solution together with tangent-space DIIS. The maximum allowed constraint-induced amplitude update was 0.1.

\section{Reference and Orbital Dependence of Tailored Coupled Cluster}

We checked how \ce{N2} TCC errors depend on the reference determinant and the orbitals used to define the active space. We tested three choices: a spin-pure CASCI(6,6) in RHF orbitals (used in the main text), the same calculation in CASSCF(6,6) natural orbitals, and an unrestricted CASCI(6,6) in broken-symmetry UHF orbitals. 

\begin{figure}[htbp]
\centering
\includegraphics[width=\linewidth]{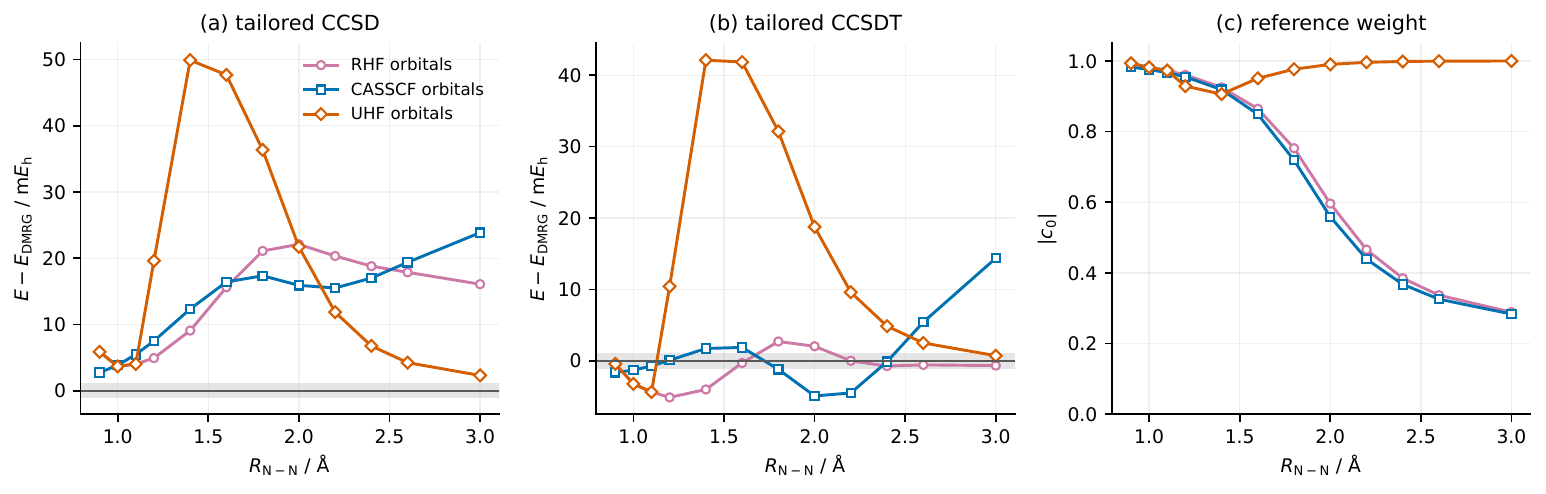}
\caption{Tailored CC for \ce{N2} in cc-pVDZ under three constructions, which differ in the reference and the orbitals the active space is built from: a spin-pure CASCI(6,6) in RHF orbitals (used in the main text), the CASSCF(6,6) natural orbitals, and an unrestricted CASCI(6,6) in broken-symmetry UHF orbitals. The references are RHF, the CASSCF natural determinant, and UHF. (a) tailored CCSD and (b) tailored CCSDT errors relative to extrapolated DMRG, with the shaded bands marking $\pm1~\mathrm{m}E_{\mathrm h}$; (c) the magnitude $|c_0|$ of the reference determinant's coefficient in the corresponding CAS wavefunction.}
\label{fig:n2_tcc_variants}
\end{figure}

Across all the bond lengths, the RHF choice gives the smallest maximum absolute error: $22.1$ and $5.1~\mathrm{m}E_{\mathrm h}$ for TCCSD and TCCSDT, compared with $23.9$ and $14.4~\mathrm{m}E_{\mathrm h}$ for CASSCF orbitals and $49.9$ and $42.1~\mathrm{m}E_{\mathrm h}$ for UHF orbitals. Its nonparallelity errors are also the smallest at both levels. The ordering can differ at individual bond lengths. For example, CASSCF-orbital TCCSD is more accurate at $2.0$~\AA.

The unrestricted results are especially poor around $1.4$--$1.8$~\AA. At $1.4$~\AA, UHF-orbital TCCSD errs by $49.9~\mathrm{m}E_{\mathrm h}$, compared with $9.1~\mathrm{m}E_{\mathrm h}$ for the RHF choice. At large separations, the unrestricted CASCI wavefunction is dominated by its reference determinant ($|c_0|\approx1$), and TCCSD approaches UCCSD. Near $1.4$--$1.6$~\AA, however, freezing the CASCI amplitudes makes TCCSD less accurate than UCCSD.